\UseRawInputEncoding
\documentclass[%
 article,
 amsmath,amssymb,
 aps,
]{revtex4-2}

\usepackage{graphicx}
\usepackage{dcolumn}
\usepackage{bm}
\usepackage{hyperref}
\usepackage[mathlines]{lineno}
\usepackage{siunitx}
\usepackage{soul}
\usepackage{xcolor}

\usepackage[textwidth=1.1\marginparwidth,backgroundcolor=yellow,linecolor=orange,textsize=scriptsize]{todonotes}

\usepackage[normalem]{ulem}

\begin{document}

\preprint{APS/PRL/ReconfigurationFramework}

\title{Simulating flow-induced reconfiguration by coupling corotational plate finite elements with a simplified pressure drag}

\author{Danick Lamoureux}
 \affiliation{
 Laboratory for Multi-Scale Mechanics (LM2), Department of Mechanical Engineering, Polytechnique Montreal, 2500 chemin de Polytechnique, Montreal (Quebec), H3T1J6, Canada
}

\author{Sophie Ramananarivo}
 
\affiliation{LadHyX, CNRS, \'Ecole Polytechnique, Institut Polytechnique de Paris, 91120, Palaiseau, France}

\author{David Melancon}
\email{david.melancon@polymtl.ca}
 \affiliation{
 Laboratory for Multi-Scale Mechanics (LM2), Department of Mechanical Engineering, Polytechnique Montreal, 2500 chemin de Polytechnique, Montreal (Quebec), H3T1J6, Canada
}
\author{Fr\'ed\'erick P. Gosselin}
\email{frederick.gosselin@polymtl.ca}
 \affiliation{
 Laboratory for Multi-Scale Mechanics (LM2), Department of Mechanical Engineering, Polytechnique Montreal, 2500 chemin de Polytechnique, Montreal (Quebec), H3T1J6, Canada
}

\date{\today}

\begin{abstract}
Developing engineering systems that rely on flow-induced reconfiguration, the phenomenon where a structure deforms under flow to reduce its drag, requires design tools that can predict the behavior of these flexible structures. Current methods include using fully coupled computational fluid dynamics and finite element analysis solvers or highly specialized theories for specific geometries. Coupled numerical methods are computationally expensive to use and non-trivial to setup, while specialized theories are difficult to generalize and take a long time to develop. A compromise between speed, accuracy, and versatility is required to be implemented into the design cycle of flexible structures under flow. This paper offers a new numerical implementation of the pressure drag in the context of a corotational finite element formulation on MATLAB. The presented software is verified against different semi-analytical theories applied to slender plates and disks cut along their radii as well as validated against experiments on kirigami sheets and draping disks.
\begin{description}
\item[Usage]
The developed code and verification cases presented here are available on GitHub \href{https://github.com/lm2-poly/FIRM}{https://github.com/lm2-poly/FIRM}.
\end{description}
\end{abstract}

\maketitle
\onecolumngrid

\section{INTRODUCTION}\label{sec:level1}

Fluid flow can cause large deformation of flexible, natural structures such as trees \citep{lin_characterizing_2023}, leaves \citep{alben_drag_2002, schouveiler_rolling_2006, gosselin_drag_2010, de_langre_scaling_2012, gosselin_mechanics_2019}, dandelions \citep{etnier_reorientation_2000, sun_drag_2023}, wings \citep{lin_bristledwings}, seaweeds \citep{martone_mechanics_2010}, and other aquatic plants \citep{zhang_flowinduced_2020}. By deforming, these biological systems reduce their drag, thereby minimizing the reaction forces required to maintain integrity \citep{gosselin_mechanics_2019}. Inspired by this natural phenomenon called flow-induced reconfiguration \citep{vogel, zhang_flowinduced_2020}, engineering applications have been proposed such as passive actuation of mechanical devices \citep{gomez_passive_2017, marzin_flow-induced_2022, marzin_origami} and energy harvesting \citep{pandey_flow-induced_2023, wang_towards_2023}. The development of these novel engineering systems rely on modelling the physics of flow-induced reconfiguration, which is currently numerically expensive or complex to develop.

Analytical and semi-analytical models have been developed to study the reconfiguration of slender structures, but they are specific to their corresponding systems, e.g., thin strips with \citep{song_three-dimensional_2022} or without flaps \citep{guttag_aeroelastic_2018}, plates with \citep{guttag_aeroelastic_2018, pezzulla_deformation_2020} or without holes \citep{gosselin_drag_2010, boukor}, disks with one cut \citep{schouveiler_rolling_2006}, multiple cuts \citep{gosselin_drag_2010} or without cuts \citep{schouveiler_flow-induced_2013}, kirigami sheets \citep{marzin_flow-induced_2022}, and dandelions \citep{etnier_reorientation_2000, sun_drag_2023}. Computational fluid dynamics (CFD) software coupled with solid mechanics finite element analysis (FEA) software have been used to predict the reconfiguration of similar slender structures \citep{hua_dynamics_2014, vanzulli_co-rotational_2023, lee_propulsion_2015, lee_flexibility_2014} and can be generalized to complex geometries. However, CFD-FEA coupled approaches are computationally expensive and complex to setup.

Hybrid solutions, defining an analytical solution for structural mechanics and using CFD to define the fluid loading, have been used \citep{pezzulla_deformation_2020}, but prove to be complex to setup.
By taking inspiration from approaches similar to \citet{taylorAnalysisSwimmingLong1952}'s semi-empirical drag model for rods set at an angle to incoming flow, analytical solutions have been derived by representing the fluid flow as a pressure field \citep{gosselin_drag_2010, schouveiler_rolling_2006, marzin_flow-induced_2022}.

Here, we propose a finite element framework based on the corotational approach that uses the pressure field obtained from a uniform flow field on a structure to model flow-induced reconfiguration of flexible structures (see Fig.~\ref{fig1}). This pressure field, which is formulated through a momentum conservation argument, has only been applied to simple geometries, typically using the Euler-Bernoulli beam equation. By implementing this fluid load within a finite element solver, we present a versatile framework, a general flow-induced reconfiguration model (FIRM), which we verify and validate on structures with increasing geometrical complexity. Our simulations successfully capture their highly nonlinear responses and match the trends predicted by semi-analytical models and experimental data reported in the literature. We conclude on a degenerate case that highlights the limitation of the FIRM framework.

\section{Methodology}
\subsection{Pressure field formulation}

When placed in a uniform flow, a thin structure, like a plate or a shell, creates a pressure difference on its windward and leeward sides, leading to pressure drag. 
To model this, we apply a similar modelling argument of momentum flux balance as \citet{schouveiler_rolling_2006}, but generalise it to more complex geometries. 
Assuming that the flow has reached a uniform permanent regime of flow velocity $\vec{U}$ (see Fig.~\ref{fig1}a-b) and averaging out in time the dynamic phenomena such as vortex shedding or turbulence, the elemental force $\text{d}\vec{F}$ produced in reaction to the flow can be derived from momentum conservation as
\begin{equation}
    \text{d}\vec{F} =  \rho_f \vec{U} \left(\vec{U}\cdot \vec{n}\right)\text{d}S,
\end{equation}
where $\rho_f$ is the fluid density. We define a Cartesian coordinate system $xyz$ aligned with the flow such that  $\vec{U}=U_\infty\vec{z}$. An infinitesimal planar element of area $\text{d}S$ of the structure has its normal $\vec{n}$ and makes an angle $\theta$  with the flow. To obtain the resulting pressure $p$ (see Fig.~\ref{fig1}c), we project $\text{d}\vec{F}$ along $\vec{n}$ and consider the force per unit area applied on this infinitesimal area
\begin{equation}\label{eq:pressurefield}
    p(\theta) = \frac{1}{2}\rho_f U_\infty^2 C_D \cos^2 \theta,
\end{equation}
where $C_D$ is the drag coefficient of the equivalent rigid structure so that the resulting drag of our reconfigurable structure can be compared to its rigid counterpart \citep{gosselin_drag_2010}. Note that in this work, we neglect skin friction as its effect is negligible on the reconfiguration of the considered structures \citep{bhatiRoleSkinFriction2018}. The drag of the structure, however, needs to be transformed back into the $xyz$ reference frame and integrated over the whole structure $\Omega$ such that 
\begin{equation}
    D = \int_\Omega 
\frac{1}{2}\rho_f U_\infty^2 C_D \cos^3\theta \text{d}\Omega.
\end{equation}

\subsection{Corotational finite element framework}

To simulate the deformation of the structure subjected to the pressure field $p(\theta)$ given by Eq. \eqref{eq:pressurefield}, we use an implementation of a corotational finite element model \citep{caselli_polar_2013}. This method decouples the rigid body motions of elements from their local deformations therefore allowing large displacements and rotations, as expected in the context of reconfiguration \citep{gosselin_drag_2010, marzin_flow-induced_2022, schouveiler_rolling_2006}. The structure is first meshed with triangular elements using GMSH \citep{GMSH} (see Fig.~\ref{fig1}d). Once meshed, the angle $\theta$ between the flow and the deformed structure is computed at each node by fitting a plane through its nearest neighboring nodes (see Fig.~\ref{fig1}e and \cite{supplementary} for more details). The deformation-dependent pressure is applied to each node and the displacements $u$, $v$, $w$ and rotations $\phi_x$, $\phi_y$, and $\phi_z$ are solved iteratively based on a finite element formulation that couples the Discrete Kirchoff Triangule (DKT) plate theory to capture bending stresses of thin plates \citep{batoz_study_1980, jeyachandrabose_alternative_1985} and the Assumed Natural Deviatoric Strain (OPT) membrane theory to capture membrane stresses \citep{felippa_study_2003} (see Fig.~\ref{fig1}f). Note that the material constitutive law is modeled as isotropic Hookean. We verify and validate the developed software using benchmark test cases (see \cite{supplementary} for more details).

\subsection{Dimensional analysis}

In the following, we use two main dimensionless numbers to characterize the reconfiguration of flexible structures: the Cauchy and reconfiguration numbers. The Cauchy number $C_Y$ is defined as the ratio of the fluid load over the elastic restoring forces 
\begin{equation}\label{eq:cauchy}
C_Y = \frac{\rho_f U_\infty^2 C_D L_c}{K},  \end{equation}
with $L_c$ a characteristic length and $K$ the structure's stiffness in the appropriate deformation mode. Low Cauchy numbers indicate either low flow velocities or stiff structures, whereas high Cauchy numbers indicate higher fluid loading.

\begin{figure}
    \centering
    \includegraphics[width=\textwidth]{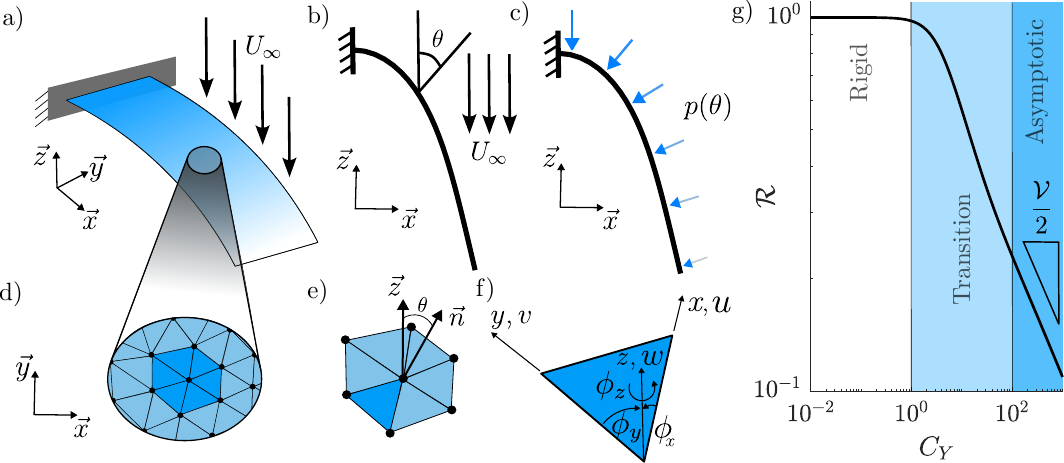}
    \caption{\textbf{Flow-Induced Reconfiguration Model (FIRM).} a)~Schematic of a flexible structure reconfiguring under a flow of velocity $U_\infty$. b)~The uniform flow field hits the structure at an angle $\theta$ that varies across the surface of the structure. c)~The flow-field is converted into a pressure field $p$ that varies as a function of the local angle $\theta$. The size of the arrows indicate the magnitude of the pressure as an example. d)~To compute its elastic deformation under flow, the structure is discretized using 2D triangular finite elements using GMSH. e)~The normal of each node $\vec{n}$ and the angle $\theta$ are calculated by fitting a plane through the nearest neighboring nodes. f)~The elements' formulation is based on a coupled Discrete Kirchoff Triangle plate theory and the Assumed Natural Deviatoric Strain membrane theory with $u$, $v$, $w$, $\phi_x$, $\phi_y$, and $\phi_z$ the displacements and rotations along the $x$, $y$, and $z$ axes. f) Typical Reconfiguration number - Cauchy number curve, showing the three regimes i) rigid behavior, ii) transition region and iii) Asymptotic behavior governed by the Vogel exponent.}
    \label{fig1}
\end{figure}

The reconfiguration number $\mathcal{R}$ relates the drag $D$ of a deformable structure with that of a rigid equivalent $D_{rigid}$
\begin{equation}\label{eq:numreconf}
     \mathcal{R} = \frac{D}{D_{rigid}} = \frac{\int_\Omega dF}{\frac{1}{2}\rho_f U_\infty^2 C_D S} = \frac{\sum_{i=1}^{N_e} \frac{1}{2}\rho_f U_\infty^2 C_D \cos^3\theta_i S_i}{\frac{1}{2}\rho_f U_\infty^2 C_D S}=\frac{\sum_{i=1}^{N_e} S_i \cos^3\theta_i}{S},
\end{equation}
where $S=\sum_{i=1}^{N_e} S_i$ is the structure's area summed over all $N_e$ elements. A reconfiguration number close to unity indicates a drag that is close to that of the rigid structure, while a lower reconfiguration number reveals that the structure has deformed sufficiently to reduce its generated drag.  Fig.~\ref{fig1}g shows a typical $\mathcal{R}$-$C_Y$ curve for a structure undergoing flow-induced reconfiguration. We see that at low Cauchy number, the reconfiguration number is close to unity in the rigid regime. As the Cauchy number is increased, a transition region first appears, followed by an asymptotic regime governed $\mathcal{R}\sim C_Y^{\mathcal{V}/2}$, where $\mathcal{V}$ is Vogel's exponent \citep{gosselin_mechanics_2019}.

\section{Results}

\subsection{Flat rectangular plate}

\begin{figure}
    \centering
    \includegraphics[width=\textwidth]{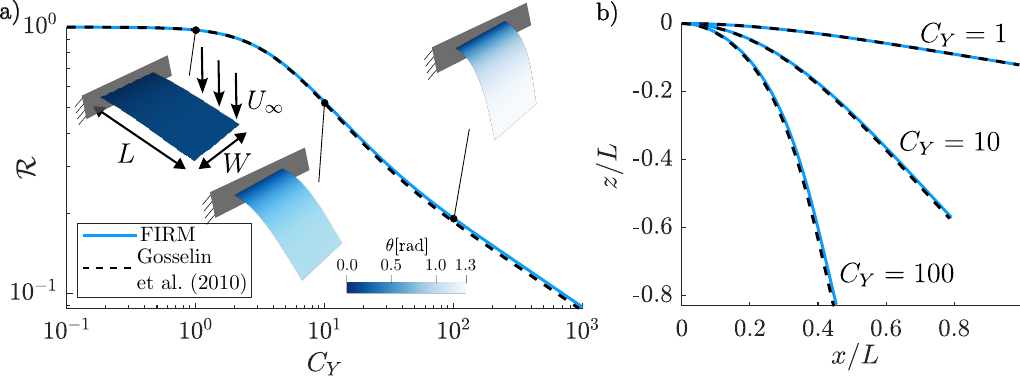} 
    \caption{\textbf{Reconfiguration of a slender plate.} a) Reconfiguration number $\mathcal{R}$ as a function of the Cauchy number $C_Y$ for a flat, slender plate of length $L$ and width $W$. The color gradient on the plate illustrates the variation of the angle of the surface's normal with the flow as shown in the colorscale. b) 2D deformed shapes of the flat rectangular plate under reconfiguration using normalized coordinates at $C_Y=\{1,10,100\}$. The dashed and solid lines represent the results from the semi-analytical model developped by Gosselin et al. (2010) and our FIRM model, respectively, as shown in a)'s legend.}
    \label{fig2}
\end{figure}

We first apply our Flow-Induced Reconfiguration Model (FIRM) to study the reconfiguration of a thin rectangular plate of width $W=\SI{35}{\mm}$, length $L=\SI{50}{\mm}$, and thickness $T=\SI{0.1}{\mm}$, clamped at one end and subjected to a flow of velocity $U_\infty$ as schematised in the inset of  Fig.~\ref{fig2}a. We model the plate material with a Young's modulus of $E = \SI{68}{\GPa}$ and Poisson's ratio of $\nu = 0$, to approach a 1D beam behavior. We consider a rigid drag coefficient  $C_D = 2.0$ and a fluid density  $\rho_f=\SI{1.225}{\kg\per\m\cubed}$. The drag coefficient in itself will not matter due to the dimensionless numbers used, but we keep a drag coefficient similar to what is found by \citet{gosselin_drag_2010}. For this case, the Cauchy number is obtained from Eq.~(\ref{eq:cauchy}) with the characteristic length $L_c = L$ and the stiffness $K=2B/L^2$ with $B=ET^3/12$ the bending stiffness. This leads to $C_Y=\rho_f U_\infty^2 C_D L^3/(2B)$. 
In order to improve solver efficiency, as the velocity is incremented, the previous solution serves as an initial guess for the next velocity iteration in a form of continuation method.

We show in Fig.~\ref{fig2}a the $\mathcal{R}$-$C_Y$ curve obtained via FIRM (solid blue line) and compare it against the results from a semi-analytical model (dashed line) based on Euler-Bernoulli beam theory that uses the same pressure formulation as in Eq. \eqref{eq:pressurefield} \citep{gosselin_drag_2010}. We find excellent agreement between the two in each of the three regimes: (i) at low Cauchy number, i.e., $C_Y < 1$, the drag of the plate is similar to its rigid counterpart ($\mathcal{R} \approx 1$), (ii) at intermediate Cauchy number, i.e., $C_Y \in [1,100]$,  there is a transition zone, and (iii) at high Cauchy number, i.e., $C_Y>100$, $\mathcal{R}$ converges to an asymptotic behavior that can be described in terms of a Vogel exponent  $\mathcal{V}=-2/3$. We show in the insets of Fig~\ref{fig2}a the deformation predicted by FIRM at $C_Y=\{1, 10, 100\}$ where the color gradient illustrates the angle the surface makes with the flow. These insets confirm that the deformation is strictly 2D. For the same Cauchy numbers, we plot in Fig.~\ref{fig2}b the deformed 2D shapes of the plates in normalized coordinates $(x/L, z/L)$ and compare the results with the semi-analytical predictions \cite{gosselin_drag_2010}. Once again, we find excellent agreement for the three deformations we observe.

\subsection{Circular plate with radial slits}

Next, we simulate with FIRM a disk with radial slits held at its center which deforms in 3D under a flow of velocity $U_\infty$ perpendicular to its undeformed shape. The disk has an external radius $R_e=\SI{3.7}{\mm}$, inner uncut radius $R_i=\SI{0.9}{\mm}$ and thickness $T=\SI{0.1}{\mm}$ cut into $N=36$ sectors, with its material modeled with a Young's modulus $E=\SI{4.848}{\GPa}$ and Poisson's ratio $\nu=0$. We consider a rigid drag coefficient $C_D=2.0$ and a flow density $\rho_f=\SI{1.225}{\kg\per\m\cubed}$. For this case, the Cauchy number is obtained from Eq.~(\ref{eq:cauchy}) with $L_c = R_e-R_i$ and $K=ET^3/\left(6\beta(R_e-R_i)^2\right)$, where $\beta=R_e/R_i$ the disk's taper ratio \citep{gosselin_drag_2010}. 

\begin{figure}
    \centering
    \includegraphics[width=\textwidth]{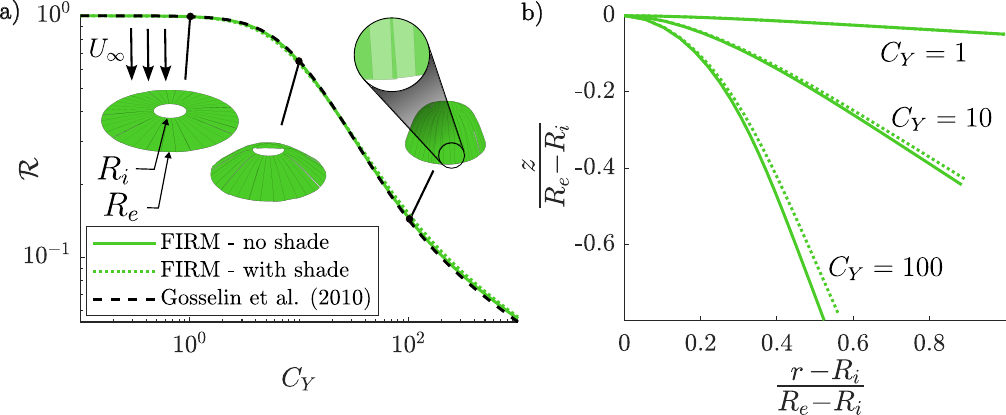}
    \caption{\textbf{Reconfiguration of a disk with radial slits.} a) Reconfiguration number $\mathcal{R}$ as a function of the Cauchy number $C_Y$ of a disk of external radius $R_e$, clamped radius $R_i$ with $N$ slits using FIRM, with and without shading, compared to ref \cite{gosselin_drag_2010}'s model. The deformed structure is illustrated in the insets at Cauchy numbers of 1, 10 and 100, with the dimensions of the disk illustrated in the first inset. The last inset's zoom shows two blades covering each other, causing shading. The current curve is produced using a taper ratio $\beta=R_e/R_i=4.111$ with $N=36$ slits. b) 2D deformed shape of a blade under shading and no shading in normalized coordinates at $C_Y = \{1,10,100\}$ as predicted by FIRM.}
    \label{fig:slitdisks}
\end{figure}

In Fig.~\ref{fig:slitdisks}a, we plot the $\mathcal{R}$-$C_Y$ curve for the described disk obtained with the FIRM software (solid green line) and compare it with the previously discussed Euler-Bernoulli beam model \citep{gosselin_drag_2010} (dashed line). There is again excellent agreement between the two models in the three previously identified regimes such that (i) for $C_Y < 1$, $\mathcal{R}\approx 1$, (ii) for $C_Y\in[1, 100]$, there is a transition region and (iii) for $C_Y > 100$, we reach an asymptotic regime that is described using a Vogel's exponent of $\mathcal{V}=-1$. The insets of Fig.~\ref{fig:slitdisks}a show the deformed shapes of the disk during these three different regimes. We note that as the blades bend under flow, they partially cover one another. This leads to shading, i.e., covered sections do not perceive the flow, and blade contact adds stiffness to the structure. In FIRM, we do not model contact and implement a simple shading law based on the exposed area of the structure (see \cite{supplementary} for more details about the implementation). We add to Fig.~\ref{fig:slitdisks}a the $\mathcal{R}$-$C_Y$ curve for this disk using the modified FIRM with the simple shading law and plot in Fig.~\ref{fig:slitdisks}b, the deformed shape of a blade at $C_Y=\{1,10,100\}$ with (solid green line) and without (broken green line) shading using its normalized radial $(r-R_i)/(R_e-R_i)$ and vertical $z/(R_e-R_i)$ position. We find that shading has minor effect on the reconfiguration of the disk with radial slits.

\subsection{Ribbon kirigami under flow}

\begin{figure}
    \centering
    \includegraphics[width=\textwidth]{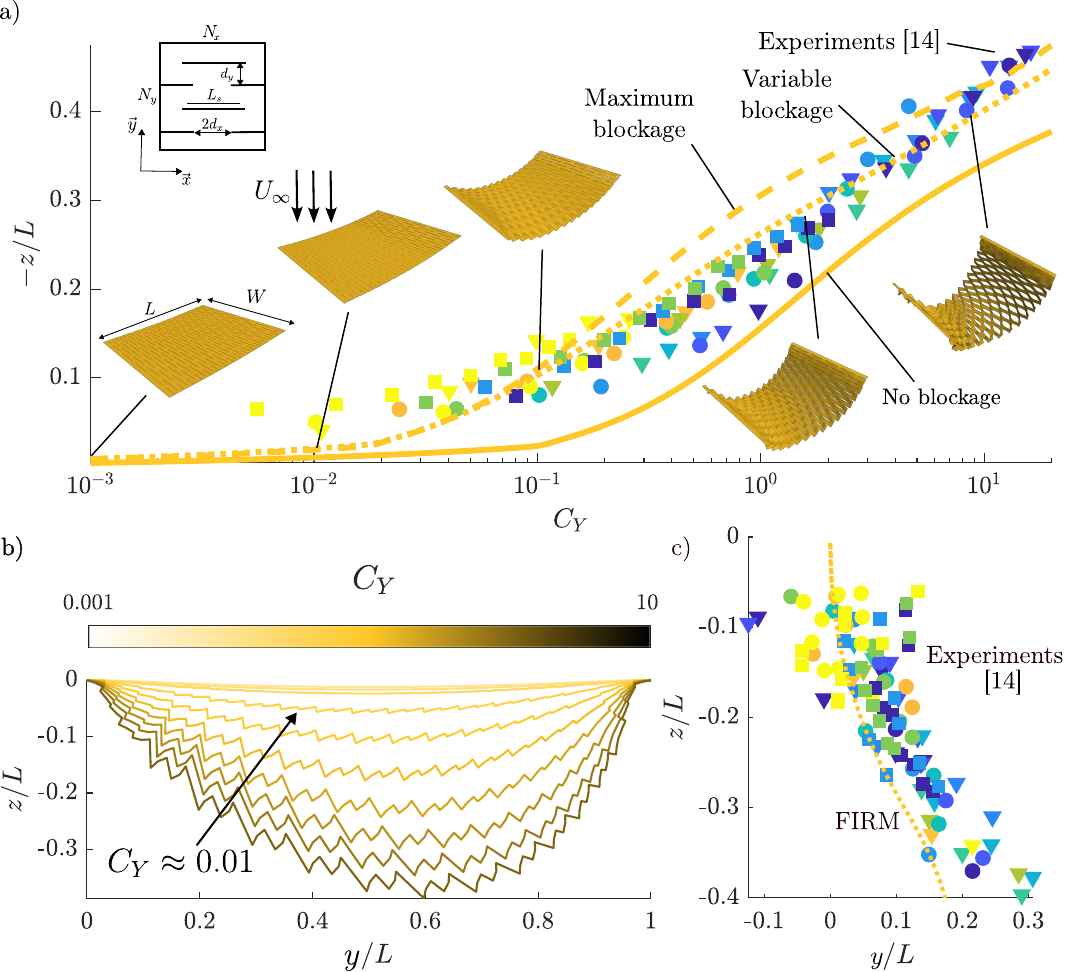}
    \caption{\textbf{Reconfiguration of a kirigami sheet under flow.} a) Dimensionless out-of-plane displacement $z/L$ as a function of the Cauchy number $C_Y$ of a flat kirigami sheet of length $L$, width $W$, and stiffness $K$ using the current implementation (FIRM), with no blockage, maximum blockage, and variable blockage assumptions compared to \citet{marzin_flow-induced_2022} experimental data. The markers and colors of the experimental data follow \citet{marzin_flow-induced_2022}'s convention once adapted to our notation, where circles are a specimen series with varied $L_s$, triangles with varied $d_y$ and squares with varied $d_x$. Insets show the sheet's deformed state at different Cauchy numbers, illustrating both in-plane stretching and post-buckling out-of-plane deformation. b) Profile of the kirigami sheet's midplane at different Cauchy numbers  in normalized deformed configuration using the variable blockage assumption. c) Evolution of the asymmetric deformation of the sheet as the out-of-plane deformation increases using the variable area velocity formulation compared to \citet{marzin_flow-induced_2022} experimental data. The markers and colors of the experimental data are the same as in a).}
    \label{fig:kirigami}
\end{figure}

To demonstrate that FIRM is able to capture complex nonlinear behaviors such as buckling, we use it to simulate the reconfiguration of a ribbon kirigami sheet under flow clamped at both ends as shown in Fig.~\ref{fig:kirigami}. We consider a sheet similar to that of \citet{marzin_flow-induced_2022} with length $L=\SI{121.6}{\mm}$, width $W=\SI{108}{\mm}$, and thickness $T=\SI{0.1}{\mm}$. The sheet is cut with $N_x=5$ slits along its width and $N_y=31$ slits along its length, as illustrated in the insets of Fig.~\ref{fig:kirigami}a. We use a notation that is different than \citet{marzin_flow-induced_2022} due to the modeling choices used during the simulation. However, we adapt the used parameters to obtain the same geometry. Slits are cut with length $L_s=\SI{17.2}{\mm}$ and pitch spaced from one another by $2d_x$ along the width of the sheet, with $d_x = \SI{2.2}{\mm}$, and $d_y=\SI{3.8}{\mm}$ along the length of the sheet.  The material is defined using a Young's modulus of $E=\SI{4}{\GPa}$ and a Poisson's ratio of $\nu=0.3$. The flowing fluid has a density $\rho_f=\SI{997}{\kg\per\m\cubed}$. Finally, we consider a rigid structure drag coefficient $C_D=2.0$.

For this case, the Cauchy number is obtained from Eq. \eqref{eq:cauchy} with $L_c = W$, and $K$ the stretching stiffness of the sheet during its second linear regime once the sheet has buckled out of plane. This stretching stiffness is evaluated numerically through a tensile simulation as $K=\SI{35.4}{\N\per\m}$ which is in good agreement with the reported value by \citet{marzin_flow-induced_2022}of $K=\SI{38.9}{\N\per\m}$ (see \cite{supplementary} for more details regarding the tensile simulation). 

Importantly, the experiments of \citet{marzin_flow-induced_2022} were conducted in a water channel of cross section $\SI{150}{\mm}\times \SI{150}{\mm}$ with an area $S_{ch}=\SI{22500}{\mm\squared}$, of which the undeformed sheet occupies  $S_0=LW=\SI{13133}{\mm\squared}$. The undeformed sheet blocks $58.4\%$ of the channel cross-section, which significantly confines the flow. As the sheet deforms under flow, its projected area is modified, changing the blockage ratio. To 
roughly estimate this effect, we use the same conservation of mass argument as \citet{marzin_flow-induced_2022} and define the effective flow velocity 
\begin{equation}
    U^* = \frac{S_{ch}}{S_{ch}-S}U_\infty,
    \label{eq:blockage}
\end{equation}
requiring the computation of the sheet frontal area   $S$ at each velocity increment.

In Fig.~\ref{fig:kirigami}a, we show the normalized out-of-plane displacement $z/L$ as a function of the Cauchy number $C_Y$ obtained via FIRM (solid yellow line) with no blockage effects ($U^*=U_\infty$) and compare it with the experimental data from \citet{marzin_flow-induced_2022}. Experimentally, the out-of-plane deformations of $19$ different kirigami sheets collapse unto a master curve that increases monotonically according to the Cauchy number. The different kirigami sheets are identified through markers indicating series (circles are specimens with varied $L_s$, triangles with varied $d_y$ and squares with varied $d_x$) and colors indicating which pattern they are precisely (we adopt the same convention as \citet{marzin_flow-induced_2022} once adapted to our notation). We notice that, for a fixed value of $C_Y$, the deformation predicted by FIRM underestimates the value of $- z/L$, which is expected as we completely neglect the increase in flow velocity due to blockage. If we consider the other extreme of maximum blockage, i.e., assuming the kigirami sheet obstructs the channel with its undeformed frontal area $A_0=LW$ for any value of $C_Y$, we find a slight overestimation of $- z/L$ shown with the yellow dashed line in Fig.~\ref{fig:kirigami}a. 
Together, the curves assuming zero and maximum blockage create an envelope that encompasses most experimental measurements. By considering a variable blockage with Eq.~\eqref{eq:blockage} at every iteration of $C_Y$, we find the yellow dotted line in Fig.~\ref{fig:kirigami}a  that lies in-between the two previous assumptions. Using this variable blockage approach, we find good agreement with the experimental data and observe the same regimes as in the experiments, i.e., (i) for $C_Y < 10^{-2}$, the displacement remains small and the kirigami sheet deforms through in-plane stretching and (ii) for $C_Y\geq 10^{-2}$, the displacement $- z/L$ suddenly increases. This sharp increase at $C_Y \approx 10^{-2}$ is due to the kirigami sheet undergoing buckling that can be highlighted in our FIRM model by plotting the profile of the sheet's midplane in Fig.~\ref{fig:kirigami}b, i.e., at $y=W/2$, for increasing Cauchy numbers. We note that this profile deforms symmetrically for low Cauchy numbers, but that at $C_Y \approx 10^{-2}$ the deformation becomes asymmetric. We quantify the predicted asymmetry in Fig.~\ref{fig:kirigami}c by plotting the maximum vertical displacement $z/L$ and its corresponding lateral $x/L$ displacement for each Cauchy number (yellow dash-dot line). Again, our prediction agrees with the experimental data from \citet{marzin_flow-induced_2022} (see markers in Fig.~\ref{fig:kirigami}c).

\subsection{Flow-induced draping of a disk}

\begin{figure}[htpb!]
    \centering
    \includegraphics[width=\textwidth]{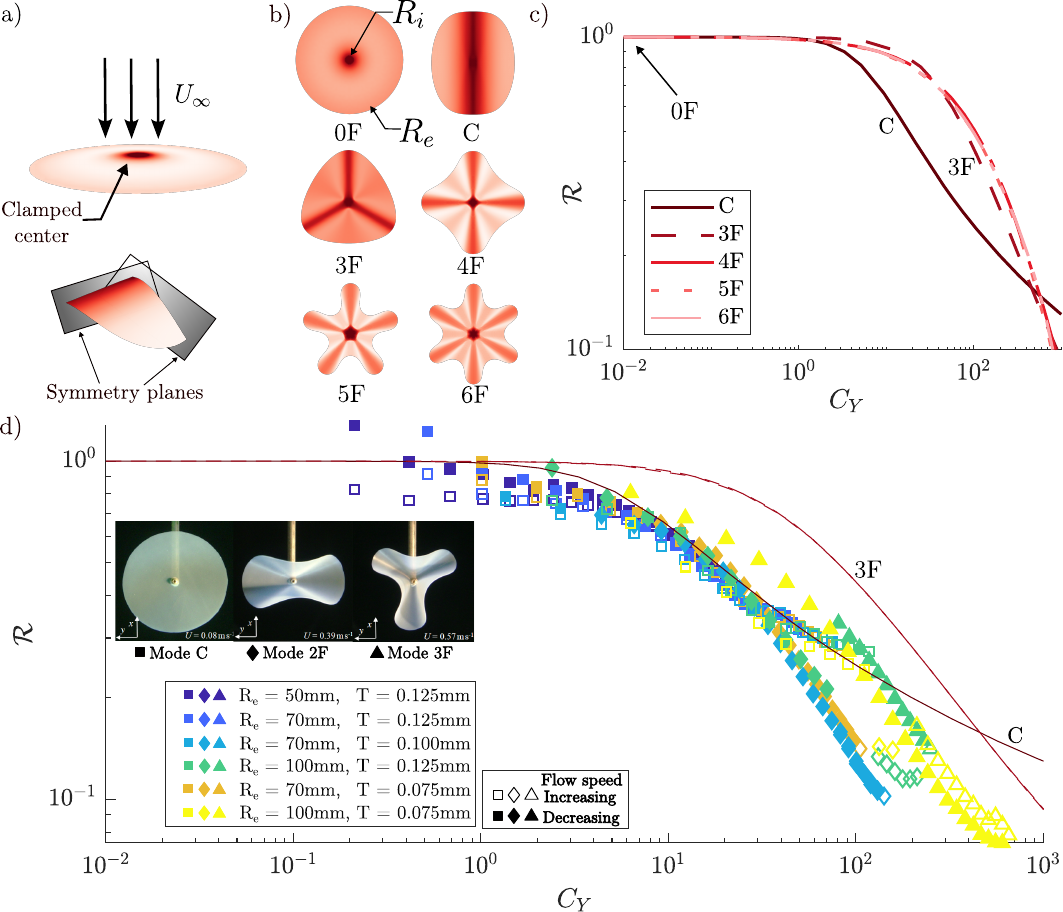}
    \caption{\textbf{Draping of a circular plate under flow.} a) Schematic of the disk under flow and its simplification through symmetry planes. b) Predicted deformation modes of a circular plate under flow. Mode $0F$ shows the external radius $R_e$ and clamped radius $R_i$ of the plate. c) Reconfiguration number $\mathcal{R}$ as a function of the Cauchy number $C_Y$ for the different deformation modes of the circular plate. d) Reconfiguration number $\mathcal{R}$ as a function of the Cauchy number $C_Y$ of modes $C$ and $3F$ compared with wind tunnel measurements. Snapshots of the experimental deformations adapted from \citet{schouveiler_flow-induced_2013} are added in the insets. \textbf{(Permission for the insets to be requested)}}
    \label{fig:disk}
\end{figure}

Finally, we test the limits of our FIRM software by studying a circular flexible plate draping, i.e., reconfiguring, under a uniform flow normal to its initial surface (see Fig.~\ref{fig:disk}a). This problem is degenerate as multiple solutions can be found for the same applied load. \citet{schouveiler_flow-induced_2013} studied experimentally the shape transition of flow-induced draping of a circular plate and observed four different modes of deformation governed by the Cauchy number: (i) a cylindrical mode (mode $C$), a (ii) two-fold (mode $2F$), and (iii) three-fold ($3F$) conical modes and (iv) a bent three-fold mode (mode $3F^*$). Importantly, to capture the transitions between each deformation mode as the flow velocity is increased, we would need an energetic approach to identify which mode would prevail over the others, which is out of the scope of this work and has already been discussed by \citet{schouveiler_flow-induced_2013}. Instead, we capture various $N$-fold symmetric deformation modes by modelling a $1/(2N)$ sector of a circular plate of radius $R_e=\SI{70}{\mm}$, clamped at its center of radius $R_i=\SI{6}{\mm}$, and of thickness $T=\SI{75}{\micro\m}$ using symmetry planes on its sides. The plate's material is modelled using a Young's modulus $E=\SI{3}{\GPa}$ and Poisson's ratio $\nu = 0.3$, with the flow modelled using a density $\rho_f=\SI{1.225}{\kg\per\m\cubed}$, the rigid drag coefficient $C_D=2.0$, and velocity $U_\infty$. For this case, the Cauchy number is obtained from Eq.~\eqref{eq:cauchy} by setting $L_c = R_e$ and $K = ET^3\ln\left(R_e/R_i\right)/\left(12(1-\nu^2)R_e^2\right)$ \cite{schouveiler_flow-induced_2013}. When we set $N=\{2,3,4,5,6\}$, we obtain the corresponding deformation modes $C$, $3F$, $4F$, $5F$ and $6F$ shown in Fig.~\ref{fig:disk}b, and mode $0F$ is obtained during all simulations for sufficiently low Cauchy numbers, as it is a stretching mode that is purely axisymmetric. In Fig.~\ref{fig:disk}b, the colormap represents the angle the disks make with the flow, but are taken at arbitrary Cauchy numbers, such that their exact value is not qualitatively important, but the colormap is used to illustrate the deformation modes. Note that when we set $N=2$, we force mode $C$ to appear, without being able to simulate mode $2F$ which coexists with mode $C$ as they are both two-fold symmetric. We are also unable to simulate mode $3F^*$ as it has only one symmetry plane and would require both contacts and shading to be implemented.

In Fig.~\ref{fig:disk}c, we show the $\mathcal{R}$-$C_Y$ curves of the simulated axisymmetric deformation modes. At low Cauchy numbers, i.e., $C_Y<1$, we find that the circular plate behaves similarly to its rigid counterpart, regardless of the imposed mode of deformation. As we increase the Cauchy number, the reconfiguration number decreases at different rates depending on the mode of reconfiguration. Below $C_Y\approx 500$, mode $C$ shows the smallest reconfiguration number, until modes $3F$ to $6F$ sequentially reach smaller reconfiguration numbers. Finally, in Fig.~\ref{fig:disk}d, we plot wind tunnel data of flow-induced draping of circular plates with varying geometry  (see \cite{supplementary} for details regarding our wind tunnel experiments) both when increasing and decreasing the flow velocity, exhibiting deformation hysteresis. The square, diamond, and triangular markers identify the mode of deformation $C$, $2F$, and $3F$, respectively. We overlay in Fig.~\ref{fig:disk}d the $\mathcal{R}$-$C_Y$ curves predicted by FIRM for modes $C$ and $3F$. We find excellent agreement with the experimental data when deforming in mode $C$. However, as mode $C$ transitions to mode $2F$ at $C_Y\approx 100$ and further transitions to mode $3F$ at $C_Y\approx 200$, there is a discrepancy between the experimental results and the numerical simulations. This discrepancy can be due to  flow physics unaccounted for by our simplified flow model, such as irregular wake phenomena. Moreover, we observe different transitions than \citet{schouveiler_flow-induced_2013}, which observed transitions from mode $C$ to $2F$ at an equivalent $C_Y=2.6$ and from $2F$ to $3F$ at an equivalent $C_Y=360$. This discrepancy with the literature could be due to plasticity effects near the creases at the center that delay mode $2F$'s onset \citep{schouveiler_flow-induced_2013}, but could also be due to dynamic and viscous effects between air and water experiments. Nevertheless, we observe that the numerical and experimental curves still follow a similar trend, they are only shifted in Cauchy.

\section{CONCLUSION AND OUTLOOK}
In conclusion, the developed FIRM framework captures complex nonlinearities observed in coupled fluid-structure interaction problems without the use of CFD while maintaining high generality and simple setup requirements. Where previous studies focused on specific theories for singular geometries or on coupled CFD-FEA software, here, we build upon a corotational finite element formulation by developing a simpler formulation of the pressure drag associated with the flow. We compare the results obtained from using this new numerical implementation with different theoretical, numerical, and experimental results obtained from the literature \cite{schouveiler_rolling_2006, gosselin_drag_2010, schouveiler_flow-induced_2013, hua_dynamics_2014, marzin_flow-induced_2022} as well as experiments performed in a wind tunnel on draping disks. Even though we do not model shading and contact physics, we find excellent agreement with many previously developed models. Moreover, it is possible to implement simpler formulations for non-modelled physics within the solver for specific cases, as was performed for the disk with radial slits and the kirigami sheet under flow. Due to the custom nature of the framework, it is also possible to add other physics on top of the flow-induced pressure field, such as buoyancy \citep{zhang_flowinduced_2020, marjoribanks_modelling_2022}, pressurized membranes \citep{windengineering}, magnetic loads \citep{abbasi_snap_2023}, etc., and therefore solve a large array of highly coupled and nonlinear problems. For example, snapping mechanisms, such as those used for flow control \citep{gomez_passive_2017, kim_flow-induced_2021} or those observed in origami structures \citep{marzin_origami}, could be studied under flow using this framework. Moreover, it is also possible to add more complex material laws into the software to model hyperelastic, visco-elastic and plastic materials.

\begin{acknowledgments}
The financial support of the Natural Sciences and Engineering Research Council of Canada (Funding Reference No. RGPIN-2019-7072) is acknowledged. D.L. acknowledges funding by a NSERC  BESC-M scholarship, the NSERC's Suppl\'ement pour \'Etudes \`a l'\'Etranger BESC-SEEMS, and a Fonds de Recherche du Qu\'ebec - Nature et Technologies's (FRQNT) master scholarship. 
\end{acknowledgments}

\bibliography{ref.bib}

\end{document}